\documentclass[twocolumn,amsmath,amssymb,10pt,superscriptaddress,a4paper,letterpaper,fleqn]{revtex4-1}
\usepackage{amssymb}
\usepackage{epsfig}
\usepackage{graphicx}
\usepackage{dcolumn}
\usepackage{array}
\usepackage{bm}
\usepackage{fancyheadings}
\usepackage{longtable}
\usepackage{multirow}
\usepackage{float}
\usepackage{afterpage}
\usepackage{color}

\setlongtables
\usepackage[breaklinks=true,linkbordercolor={1 1 1}]{hyperref}

\begin{document}
\setcounter{page}{1}

\title{
\qquad \\ \qquad \\ \qquad \\  \qquad \\  \qquad \\ \qquad \\
Cross Sections of Neutron Reactions $(n,~p)$, $(n,~\alpha )$, $(n,~2n)$ on Isotopes  of
Dysprosium,  Erbium and  Ytterbium at $\sim$ 14 MeV Neutron Energy}

\author{A.O. Kadenko}
\email[Corresponding author, electronic address:\\ ]{kadenkoartem@gmail.com}
\affiliation{Nuclear Physics Department, Taras Shevchenko National University, Kyiv, Ukraine}

\author{I.M. Kadenko}
\affiliation{Nuclear Physics Department, Taras Shevchenko National University, Kyiv, Ukraine}

\author{V.A. Plujko}
\affiliation{Nuclear Physics Department, Taras Shevchenko National University, Kyiv, Ukraine}

\author{O.M. Gorbachenko}
\affiliation{Nuclear Physics Department, Taras Shevchenko National University, Kyiv, Ukraine}

\author{G.I. Primenko}
\affiliation{Nuclear Physics Department, Taras Shevchenko National University, Kyiv, Ukraine}

\date{\today}

\begin{abstract}
{
The cross sections of the nuclear reactions induced by neutrons at $E_n$= 14.6~MeV on the isotopes of Dy, Er, Yb with emission of neutrons, proton and alpha-particle are studied by the use of new experimental data and different theoretical approaches. New and improved experimental data are measured by the neutron-activation technique. The experimental  and evaluated  data from EXFOR, TENDL, ENDF libraries are compared with different systematics and calculations by codes of EMPIRE~3.0 and TALYS~1.2. Contribution of pre-equilibrium decay is discussed.  Different systematics for estimations of the cross-sections of considered nuclear reactions are tested.
}
\end{abstract}
\maketitle


\lhead{Cross Sections of Neutron Reactions $\dots$}
\chead{NUCLEAR DATA SHEETS}
\rhead{A.O. Kadenko \textit{et al.}}
\lfoot{}
\rfoot{}
\renewcommand{\footrulewidth}{0.4pt}

\begin{sloppypar}

\section{ INTRODUCTION}

Studies of the nuclear reaction cross sections induced by neutrons provide information on the properties of excited states of atomic nuclei and nuclear reaction mechanisms \cite{09Capo}. Data on nuclear reaction cross section are also needed in  applications, specifically, such as the design of fusion reactor protection and modernization of existing nuclear power plants \cite{06Forr,09Koni}.
Despite of a large amount of information \cite{12EXFO} on observed characteristics of neutron interactions with nuclei, there are disagreements between existing experimental data and evaluated data both within different systematics and calculations by the different codes.

In this contribution, experimental and theoretical  cross sections of reactions (n, p), (n, $\alpha $), (n, 2n) on Dy, Er and Yb isotopes at neutron energies near 14.6 MeV are determined and compared.
Neutron generator (NG-300), installed in Nuclear Physics Department  of Taras Shevchenko National University of Kyiv, was used
as a source of neutrons with $E_n$= 14.6 $\pm$ 0.2 MeV. The neutron activation method was applied for measurements of the cross-sections(see \cite{12Gorb,12Kade} for details).
Theoretical calculations were performed by the EMPIRE 3.0  and TALYS 1.2 codes\cite{07Herm,08Koni}. The experimental cross-sections were also compared with data from the latest versions of evaluated nuclear data libraries: ENDF/B-VII, TENDL-2010, JENDL-4.0. The reliability of the different systematics \cite{71LuWe}-\cite{09Kono} for estimation of the nuclear reaction cross sections on isotopes of Dy, Er and Yb was  analyzed.

\section{RESULTS OF MEASUREMENTS AND CALCULATIONS }


\begin{figure}[!htb]
\includegraphics[width=0.95\columnwidth]{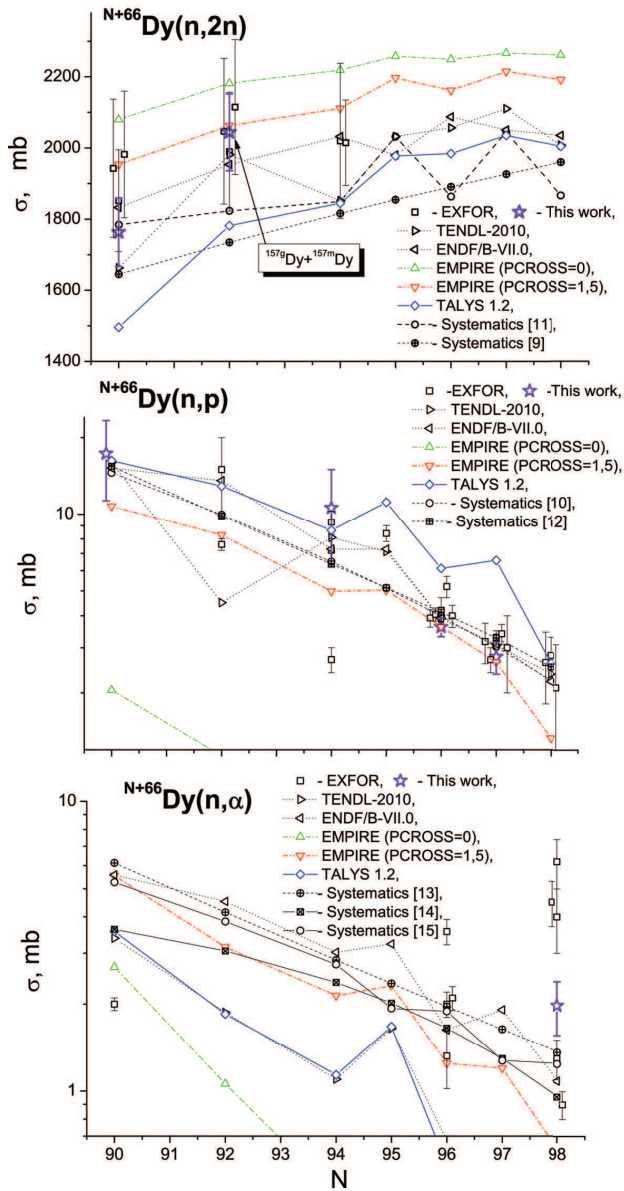}
\caption{Cross sections of the (n, 2n), (n, p) and (n, $\alpha $) reactions  on the isotopes of Dysprosium.}
\label{fig1}
\end{figure}

Figure \ref{fig1} shows cross sections of the (n, 2n), (n, p), (n, $\alpha$)  reactions  on the isotopes of Dy
as a function of the number of neutrons in comparison with theoretical calculations and evaluated data.
For reactions with transitions on ground (g) and metastable (m) states, the residual nuclei are additionally indicated on this and next figures.

Calculations of the nuclear reaction cross sections by the EMPIRE~3.0 code were performed with and without including pre-equilibrium processes (PCROSS~=~1.5 and ~0). For nuclear level density, the generalized superfluid model was
taken. The global optical potential of Koning-Delaroche \cite{03Koni} was used. In calculations by the TALYS 1.2 code
default parameters were set.  It can be seen, that  the measured  (n,2n) cross section on the  $^{158}$Dy (N=92) with allowance for uncertainties coincides with the available experimental data.  The measured (n,2n) cross section on the $^{156}$Dy (N=90) is less than previous ones, but it coincides with evaluated value from the TENDL-2010 library.  The (n, 2n) cross sections  on Dy isotopes increase with the  neutron number increasing.

The allowance for pre-equilibrium processes  leads to strong increasing the (n, p) cross sections (approximately in five times). A behavior of the (n, 2n) cross sections is opposite, and  pre-equilibrium emission reduce the values of these cross sections  mainly due to increase cross-sections of  competing binary reactions with emission of charge particles.

\begin{figure}[!htb]
\includegraphics[width=0.95\columnwidth]{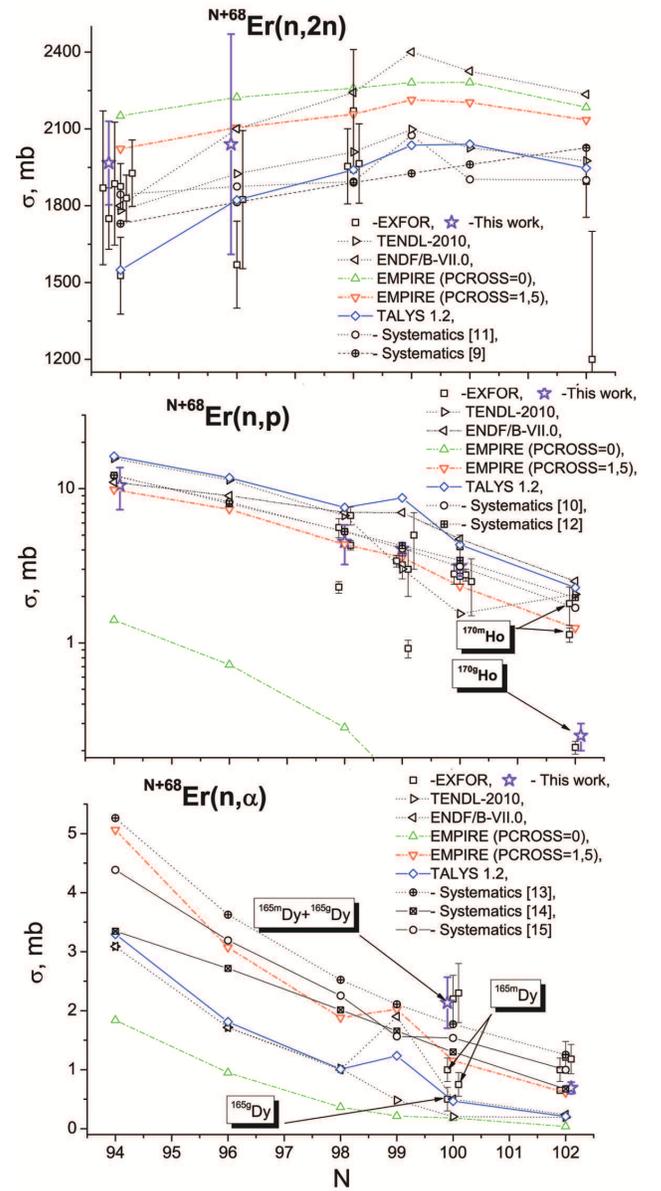}
\caption{Cross sections of the (n, 2n), (n, p) and (n, $\alpha $) reactions  on the isotopes of Erbium.}
\label{fig2}
\end{figure}

On the whole,  the results of calculations by EMPIRE 3.0 code with allowance for pre-equilibrium processes
are in best agreement with experimental data. For calculations using different systematics,
the results according to  \cite{99Kono} are more suited for description of the experimental data.

Cross sections of the (n, 2n), (n, p), (n, $\alpha$)  reactions  on the Er isotopes are given on figure \ref{fig2}.
Cross sections of (n, 2n) reaction on the Er isotopes have the similar peculiarities as on Dy.
There is a rather good agreement between presented measurements and the results of other authors.

\begin{figure}[!htb]
\includegraphics[width=0.95\columnwidth]{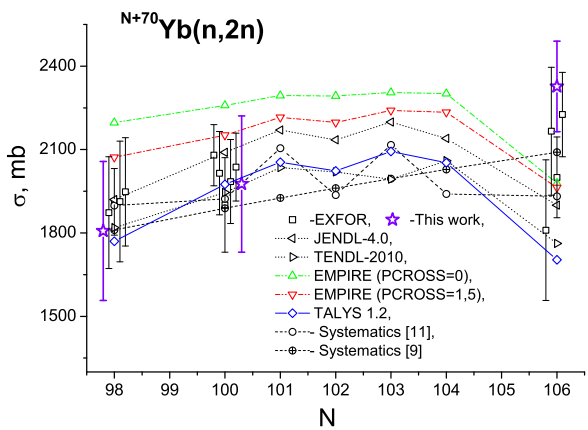}
\caption{Cross sections of the (n, 2n) reaction  on the isotopes of  Ytterbium.}
\label{fig3}
\end{figure}

\begin{figure}[!htb]
\includegraphics[width=0.95\columnwidth]{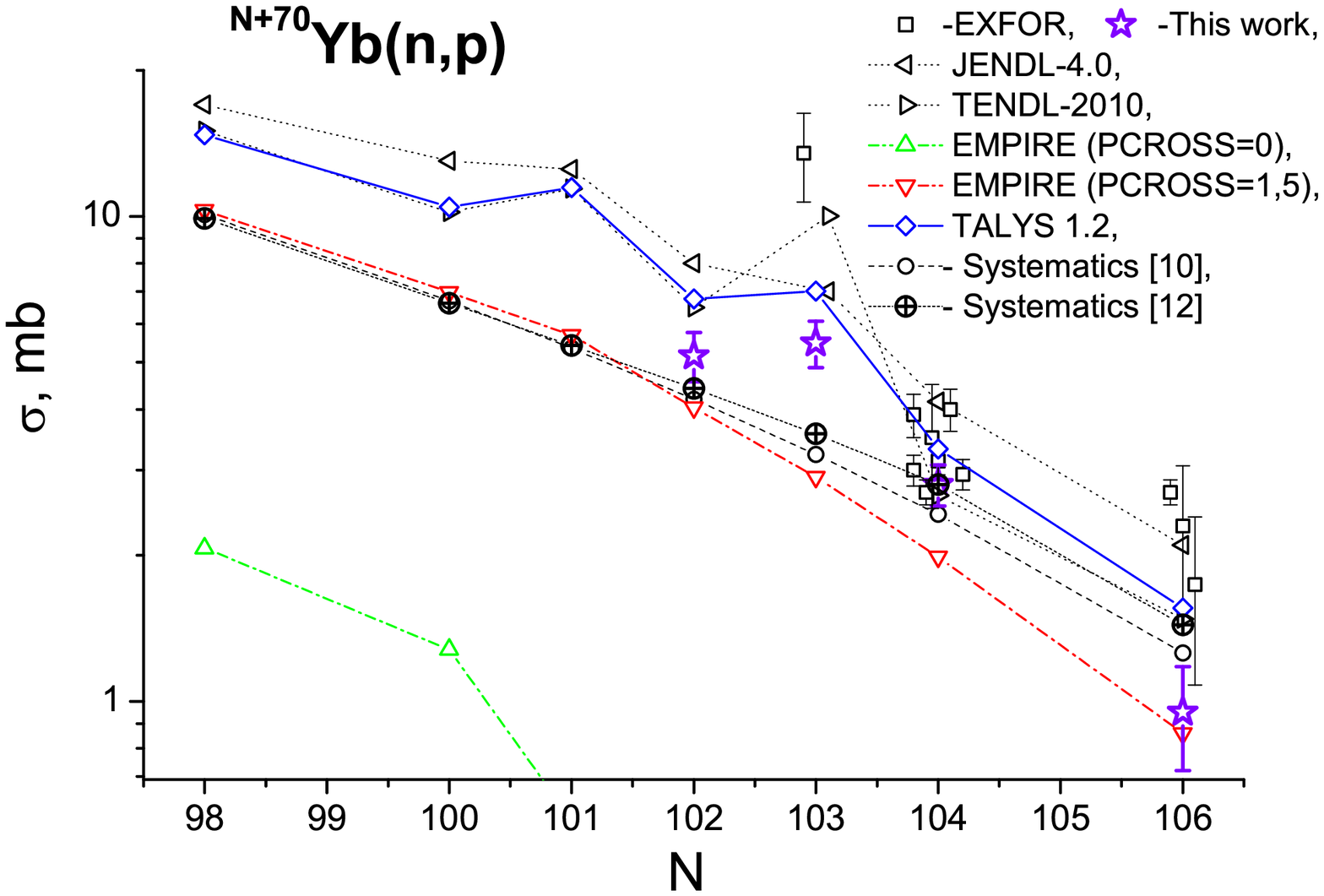}
\caption{Cross sections of the (n, p)  reaction  on the isotopes of  Ytterbium.}
\label{fig4}
\end{figure}

\begin{figure}[!htb]
\includegraphics[width=0.95\columnwidth]{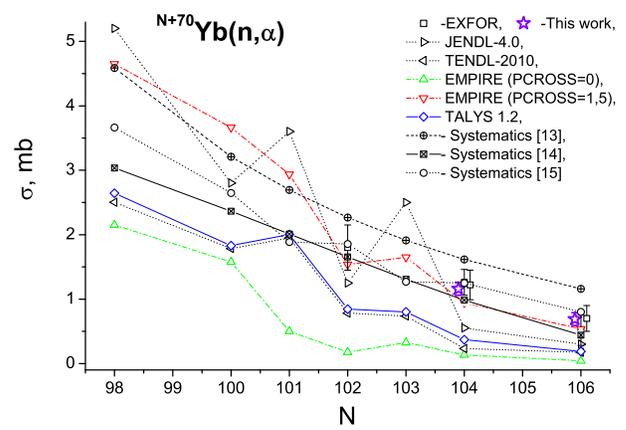}
\caption{Cross sections of the (n, $\alpha $) reaction  on the isotopes of  Ytterbium.}
\label{fig5}
\end{figure}

Figures \ref{fig3}-\ref{fig5}
show cross section of the (n, 2n), (n, p), (n, $\alpha$)  reactions  the isotopes of Yb.
Rather good agreement between experimental data for the $^{168}$Yb and $^{170}$Yb isotopes (N = 98 and 100) is observed for (n,2n) reaction, but  presented  cross section on $^{176}$Yb (N = 106) is placed higher.
The cross section of the reaction $^{172}$Yb(n,~p)$^{172}$Tm (N=102) calculated by the EMPIRE 3.0 with pre-equilibrium processes
is agree better with measured value.  The results of calculation using systematics
from \cite{06Broe,05Belg} is also closer to measured one.
The  cross sections of the (n,$\alpha$) reaction were measured with higher precision and they agree better with
calculation by systematics from \cite{08Kade,09Kono}.

\vspace{0.5cm}

\section{CONCLUSIONS}

The results of measurements of the cross sections of the nuclear reactions (n, p), (n, $\alpha $), (n, 2n)  on isotopes of Dy, Er and Yb at the neutron energy 14.6 $\pm$ 0.2 MeV are presented. They were compared with available experimental data, evaluated nuclear data and the theoretical calculations by the EMPIRE and TALYS codes. In the most cases, the presented  data   correlate well with available experimental data. On the whole,  the cross sections calculated by the  EMPIRE 3.0 code with  pre-equilibrium processes  agree better with experimental data than the results obtained by  TALYS 1.2 code with default set of parameters. Amongst the systematics, the cross section values calculated by expressions from \cite{06Broe,05Belg} are more consistent with measured cross-sections.
\end{sloppypar}

\end{document}